\documentstyle[aps,epsf,prd,eqsecnum]{revtex}

\begin{document}
%
\vskip8mm
\begin{center}{\large
\bf Bulk scalar field in the braneworld can mimic the 4d inflaton dynamics
 }
\end{center}
\vspace*{4mm}
\centerline{\large
Yoshiaki Himemoto\footnote{E-mail:himemoto@vega.ess.sci.osaka-u.ac.jp}
Takahiro Tanaka\footnote{E-mail:tanaka@yukawa.kyoto-u.ac.jp}\kern-0.3ex${}^{\star}$
and Misao Sasaki\footnote{E-mail:misao@vega.ess.sci.osaka-u.ac.jp}}

\vspace*{4mm}
\centerline{\em Department of Earth and Space Science, Graduate
School of Science}
\centerline{\em Osaka University, Toyonaka 560-0043, Japan;}

\centerline{\em ${}^{\star}$Yukawa Institute for Theoretical Physics,
Kyoto University, Kyoto 606-8502, Japan}
\begin{abstract}
Based on the recently proposed scenario of
inflation driven by a bulk scalar field in the
braneworld of the Randall-Sundrum (RS) type,
we investigate the dynamics of a bulk scalar field on
the inflating braneworld.
We derive the late-time behavior of the bulk scalar field
by analyzing the property of the retarded Green function.
We find that the late-time behavior is basically dominated by
a single (or a pair of) pole(s) in the Green function
irrespective of the initial condition
and of the signature of $m^{2}=V''(\phi)$, where $V(\phi)$
is the potential of the bulk scalar field.
Including the lowest order backreaction to the geometry,
this late-time behavior can be well approximated
by an effective four-dimensional scalar field with $m^2_{\mathrm{eff}}=m^2/2$.
The mapping to the four-dimensional effective theory
is given by a simple scaling of the potential with
a redefinition of the field.
Our result supports the picture that
the scenario of inflation driven by a bulk scalar field
works in quite a similar way to that in the standard
four-dimensional cosmology.
\end{abstract}

\vskip10mm

\section{Introduction}

It is very likely that our four-dimensional universe
is a subspace in a higher-dimensional spacetime.
In fact, string theory, which is a candidate for the unified theory,
is a higher-dimensional theory.
As a realization of such a higher-dimensional theory,
the braneworld scenario has attracted a lot of attention recently.
The new idea of the braneworld is that the matter fields
are confined on the brane, while the gravitational
fields propagate in higher dimensions which we call the bulk.
Based on M-theory, Horava and Witten showed the possibility that
the matter fields may consistently appear on the 10-dimensional
boundary of a $Z_2$-symmetric 11-dimensional
spacetime\cite{Horava:1996qa}.
Using the fact that only gravitons propagate through the bulk,
Arkani-Hamed, Dimopoulos and Dvali
proposed a new solution to the hierarchy problem between the
Planck and electroweak scales\cite{Arkani-Hamed}.
Then Randall and Sundrum (RS) proposed a model of a braneworld
in which the bulk is a five-dimensional anti-de Sitter (AdS$_5$) space
with the brane(s) as the fixed point(s) of the
$Z_2$ symmetry\cite{Randall:1999ee,Randall:1999vf}.
In particular, the second model proposed by Randall and
Sundrum\cite{Randall:1999vf}
has a single brane, hence the extension in the
direction of the extra dimension is infinite.
Nevertheless, the spacetime is effectively compactified
within the curvature radius $\ell$ of AdS$_5$,
and the four-dimensional Einstein gravity is effectively recovered on
the brane for length scales greater than
$\ell$\cite{Randall:1999vf,Garriga:2000yh}.
Thus this second RS braneworld model provides a unique mechanism
for compactification called warped compactification.
Because of this attractive feature,
a lot of work has been done on the cosmological
aspect of the model to see if there is any significant
difference from the standard four-dimensional
cosmology\cite{cosmology,bulky,Shiromizu:2000wj,Garriga:2000bq}.

It is natural that the five-dimensional action includes
some scalar fields of gravitational origin
from the viewpoint of unified theories in a yet higher-dimensional
spacetime.
Braneworld models with such a scalar field have been discussed by
various authors\cite{hawking}.
Then one can consider a scenario in which
inflation on the brane is driven by
a scalar field living in the bulk in place of
an inflaton field confined on the brane\cite{Himemoto:2001nd}.
Provided there is a bulk scalar field with a suitable potential,
it was shown in Ref.~\cite{Himemoto:2001nd}
that there exists a field configuration in the bulk
that indeed realizes inflation on the brane.
Subsequently, the quantum fluctuations in this scenario of
brane inflation were
investigated \cite{Kobayashi:2001yh,Sago:2001gi}.
In particular, it was
shown that the correction due to the five-dimensional nature
of the inflaton field is small,
as long as $|m^{2}|\ell^{2}\ll 1$
and $H^{2}\ell^{2}\ll 1$ are satisfied,
where $\ell$ is the curvature radius of AdS$_5$,
$m$ is the mass of the bulk scalar
field, and $H$ is the Hubble parameter on the brane
\cite{Sago:2001gi}.
Furthermore, the reheating after inflation
based on this scenario has been discussed by Yokoyama and
Himemoto\cite{Yokoyama:2001nw},
and they found that the standard inflationary cosmology is
reproduced under the same conditions.

However, these results for this new scenario are
more or less based on a particular scalar field
configuration discussed in \cite{Himemoto:2001nd},
the details of which will be explained in the next section.
Therefore, strictly speaking, the viability and generality
of this scenario should be more carefully examined.
To overcome the limitations of the previous analyses,
in this paper we develop a method to
discuss the dynamics of a bulk scalar field
in more general situations, which is
applicable to most of the stages of inflation
including the reheating era.

This paper is organized as follows.
In Sec.~II, we review the inflation scenario driven by
a bulk scalar field. In Sec.~III,
we extract the behavior of the bulk scalar field
on the inflating braneworld from the properties of the Green function.
In Sec.~IV, we discuss the effective description of
the dynamics from the four-dimensional point of view
and show that there is a simple correspondence
between the 5-dimensional potential $V(\phi)$ and
the effective four-dimensional potential $V_{\mathrm{eff}}(\Phi)$
of the induced scalar field $\Phi$ on the brane.
Section V is devoted to discussions.

\section{Braneworld inflation without inflaton on the brane}

We consider a five-dimensional bulk
with a single positive tension brane. The brane is located
at the fixed point of the $Z_2$ symmetry.
We adopt the Gaussian normal coordinates and express
the metric as
\begin{equation}
ds^{2}  = g_{a b} dx^{a}dx^{b}=dr^{2}
+q_{\mu \nu}dx^{\mu}dx^{\nu}\,,
\label{gmetric}
\end{equation}
where the brane is assumed to be located at $r=r_0$.
As for degrees of freedom which propagate in the bulk,
we consider a scalar field as well as gravitons.
We denote the energy momentum tensor confined on the brane
by $S_{a b}$ and that spread over the bulk by $T_{ab}$.
We introduce the five-dimensional negative
cosmological constant $\Lambda_{5}$ as an offset of the
vacuum energy apart from $T_{ab}$.

The five-dimensional Einstein equations are
\begin{equation}
  R_{ab}-{1\over2}g_{ab}R+\Lambda_5g_{ab}
=\kappa_5^2\left(T_{ab}+S_{a b} \delta(r-r_0)\right)\,.
\label{bulkgreq}
\end{equation}
As for the forms $S_{ab}$ and $T_{ab}$,
we neglect the contribution to $S_{ab}$ from
the matter fields confined on the brane
and consider a minimally coupled bulk scalar field
with the potential $V(\phi)$.
Thus we have
\begin{eqnarray}
S_{a b} &=& -\sigma q_{a b}, \cr
T_{a b} &=& \phi_{,a} \phi_{,b}- g_{a b}\left( {1\over2}
g^{c d}\phi_{,c} \phi_{,d}+ V(\phi)\right),
\label{5emt}
\end{eqnarray}
where $\sigma$ is the tension of the brane.
For simplicity, we do not consider a possible
coupling of $\phi$ to the brane
metric $q_{\mu\nu}(r_0)$, though an extension
to such a case may be worth investigating in the future.
In order to recover the Randall-Sundrum flat braneworld
when $T_{ab}$ vanishes,
we choose $\Lambda_5$ as
\begin{equation}
\Lambda_5=-{\kappa_5^4\sigma^2\over 6}.
\label{tension}
\end{equation}
Then, the effective four-dimensional Einstein equations on the brane
become
\cite{Shiromizu:2000wj,Himemoto:2001nd}
\begin{equation}
G_{\mu \nu}=
\kappa_{4}^{2}T_{\mu \nu}^{(s)}-E_{\mu \nu},
\label{Einstein}
\end{equation}
where
\begin{eqnarray}
\kappa_{4}^{2}&=&{\kappa_{5}^{4}\sigma \over 6}, \\
T_{\mu \nu}^{(s)} & = &{1\over{\kappa_5^{2} \sigma}}
\left(4 \phi_{,\mu} \phi_{,\nu}+\left({3\over2}(\phi_{,r})^{2}
-{5\over2}q^{\alpha \beta}\phi_{,\alpha} \phi_{,\beta}
-3V(\phi)\right)
q_{\mu \nu}\right),\\
E_{\mu \nu}&=& \,{}^{(5)}C_{r b r d}
\,q_{\mu}^{b}\,q_{\nu}^{d}\,.
\end{eqnarray}
Here, ${}^{(5)}C_{r b r d}$
is the five-dimensional Weyl tensor with its two indices
projected in the $r$ direction.

Focusing on the zeroth-order description of the cosmological model,
we consider the case in which the metric induced on the
brane is isotropic and homogeneous,
\begin{equation}
\left.ds^{2}\right|_{r=r_0}
=q_{\mu\nu}(r=r_0)dx^\mu dx^\nu
=-dt^{2}+a(t)^{2}\gamma_{i j}dx^{i} dx^{j},
\end{equation}
where $\gamma_{ij}$ is the metric of a constant curvature space.
Because of the assumed $Z_2$ symmetry, the boundary condition for
the bulk scalar field at the position
of the brane is given by
\begin{equation}
\partial_{r}\phi|_{r=r_0}=0.
\label{bc}
\end{equation}
Here we have further assumed that the bulk scalar does not
couple to the metric on the brane.
Then, the time-time component of Eq.~(\ref{Einstein})
gives the four-dimensional effective Friedmann equation
\begin{equation}
3\left[\left({\dot a\over a}\right)^{2}+{K\over a^2}\right]
\equiv 3H^{2}
=\kappa_4^{2}\rho_{\mathrm{eff}},
\label{friedmann}
\end{equation}
with
\begin{eqnarray}
\rho_{\mathrm{eff}}={3\over{\kappa_5^{2}\sigma}}
\left({\dot{\phi}^{2}\over2}+V(\phi)\right)
-{E_{tt}\over\kappa_4^{2}}.
\label{rhoeff}
\end{eqnarray}
The equations for $\phi$ and $E_{tt}$,
are basically five-dimensional.
However, in the present spatially homogeneous case,
the Bianchi identities supply the evolution equation of
$E_{tt}$ on the brane as\cite{Himemoto:2001nd}
\begin{equation}
E_{tt}={\kappa_5^{2}\over{2a^{4}}}\int^t a^{4}\dot{\phi}
(\partial_{r}^{2}\phi + {\dot a\over a}\dot{\phi})\,dt.
\label{weyl}
\end{equation}
Hence, once we obtain $\phi$ by solving the five-dimensional equation,
the value of $E_{tt}$ on the brane is derived from
this equation without solving the five-dimensional one.

In the previous work\cite{Himemoto:2001nd},
we considered the five-dimensional evolution of a bulk scalar field
with the potential $V(\phi)$ in the slow-roll limit.
We approximated the energy density in the bulk
by a fixed value of the cosmological constant
\begin{equation}
\Lambda_{5,\mathrm{eff}}=\Lambda_{5}+\kappa_5^{2}V(\phi_{0}),
\end{equation}
where $\phi_{0}$ is an appropriate representative
value of $\phi$.
Then, neglecting the $\dot\phi^2$ and $E_{tt}$ terms in the
effective energy density, the effective curvature radius $\ell$ and
the Hubble constant $H$ are given by
\begin{equation}
\ell^{2}={6\over|\Lambda_{5,\mathrm{eff}}|}\,, \quad
H^{2}={\kappa_5^{2}\over6}V(\phi_{0}),
\label{effective}
\end{equation}
and the bulk metric takes the form
\begin{eqnarray}
ds^2=dr^2+(H\ell)^2\sinh^2(r/\ell)(-dt^2+a^2(t)\gamma_{ij}dx^i dx^j).
\label{bulkmetric}
\end{eqnarray}
Note that
\begin{equation}
  H\ell={1\over \sinh(r_0/\ell)}\,.
\end{equation}
As a toy model, we considered the potential of the form
\begin{equation}
 V(\phi)=V_{0}+{1\over 2}m^{2}\phi^{2}.
\label{potential}
\end{equation}
Under these assumptions,
we found that there is a unique nodeless solution
in the separable form as
\begin{eqnarray}
\phi=u(r)\psi(t).
\label{sepform}
\end{eqnarray}
When $|m^{2}\ell^{2}| \ll 1$ and $H^{2}\ell^{2} \ll 1$,
the time dependence of the solution turns out to
be identical to that of a four-dimensional scalar with the
effective mass squared given by $m^2_{\mathrm{eff}} = m^2/2$.
Further, substituting this solution into the formula for $E_{tt}$,
the effective energy density $\rho_{\mathrm{eff}}$
turns out to be identical to that for the
effective four-dimensional scalar field defined by
$\Phi =(6/|\Lambda_5|)^{1/4}\phi(r_0)$ whose mass squared is
$m^2_{\mathrm{eff}}$.
This fact indicates that the dynamics of this system
may be well described by the effective
four-dimensional scalar field $\Phi$
when both $|m|\ell$ and $H\ell$ are small,
and the standard slow-roll inflation is realized
by the bulk scalar field.
\footnote{Slow-roll inflation is realized
on the brane if $|m^{2}|/H^{2} \ll 1$, even if
$|m^{2}|\ell^{2} \gg 1$ and $H^{2}\ell^{2} \gg 1$.
However, the effective energy density has a small discrepancy
from the standard one in this case\cite{Himemoto:2001nd}.}

However, since the above solution is obtained by imposing the
constraint of separability, it is not clear to what extent
we can trust the speculations based on this specific solution.
Moreover, the analysis in Ref.~\cite{Himemoto:2001nd} was
restricted to the case with $m^{2}<0$
to avoid the singular behavior of the solution at $r=0$.
In this paper, we give a rigorous
foundation of the arguments given in
the previous papers, and show that the scenario is valid
for a much more general class of initial conditions
for which there is no need to assume
the separable form (\ref{sepform}) nor the case $m^2<0$.

\section{Bulk scalar dynamics}
We consider the evolution of a bulk scalar field on the
background spacetime
which consists of the AdS$_5$ bulk and the boundary
de Sitter brane. We consider arbitrary, regular initial data for this
scalar field and investigate the generic behavior of the scalar
field at sufficiently late-times by
analyzing the properties of the retarded Green function.

\subsection{Construction of the Green function}
The bulk geometry is given by Eq.~(\ref{bulkmetric})
with $a(t)$ of the de Sitter space.
For simplicity, we take the spatially flat chart
of the de Sitter space. Then the metric becomes
\begin{equation}
 ds^2=dr^2+(H \ell)^2\sinh^2(r/\ell)\left[
    -dt^2+H^{-2}e^{2Ht} d{\bbox{x}}^2_{(3)}\right].
\label{desitter}
\end{equation}
The field equation for $\phi$ is
\begin{equation}
(-\Box_{5}+m^{2})\phi=
  \left[-\hat L_r +{1
    \over (H\ell)^2 \sinh^2(r/\ell)}\left(
  \hat L_t-H^2 e^{-2Ht}\partial_{\bbox{x}}^2 \right)\right]
  \phi=0,
\label{fieldequ}
\end{equation}
where
\begin{equation}
\hat L_t={\partial^2\over \partial t^2}+3H{\partial\over\partial t},
\end{equation}
and
\begin{equation}
\hat L_r={1\over \sinh^4(r/\ell)}{\partial\over\partial r}
\sinh^4(r/\ell){\partial\over\partial r} -m^2.
\label{spaceeq}
\end{equation}
The retarded Green function satisfies
\begin{equation}
(-\Box_{5}+m^{2})G(x,x')={\delta^{5}(x-x')\over \sqrt{-g}}\,,
\label{green}
\end{equation}
with the causal condition that $G(x,x')=0$ for $x'$ not in the
causal past of $x$.
For given initial data on the hypersurface $t=t_i$,
the time evolution of a scalar field is given by
\begin{eqnarray}
 \phi(x)
= \int_{t'=t_i}
           [(n^{a}\partial'_{a} G(x,x'))\phi(x')
-G(x,x') n^{a}\partial'_{a} \phi(x')]
            \sqrt{\gamma(x')}\  d^{4}x',
\end{eqnarray}
where $n^{a}$ is the timelike unit vector normal
to the initial hypersurface
and $\gamma$ is the determinant of the metric induced on
this initial hypersurface.

Since we are interested in the spatially homogeneous brane,
we focus on the $\bbox{x}$-independent scalar field
configurations.
Namely, we consider the spatially averaged Green function
defined by
\begin{equation}
{\cal G}(t,r;t',r'):=\int d{\bbox{x}}'\, G(x,x').
\end{equation}
Since $G(x,x')$ depends on the spatial
coordinates ${\bbox{x}}$ and ${\bbox{x}}'$ though
the form $|{\bbox{x}}-{\bbox{x}}'|$,
the ${\bbox{x}}$ dependence also
disappears after taking the average over ${\bbox{x}}'$.
The equation for ${\cal G}(t,r;t',r')$ follows
from Eq.~(\ref{green}) as
\begin{equation}
  \left[{\hat L_t\over (H\ell)^2\sinh^2(r/\ell)}-\hat L_r
   \right]{\cal G}(t,r;t',r')={\delta(t-t')\delta(r-r')
    \over H\ell^4\sinh^4(r/\ell)e^{3Ht}}\,.
\label{green2}
\end{equation}

Let us now construct the Green function ${\cal G}(t,r;t',r')$.
We begin by considering a set of eigenfunctions of
the operator $\hat L_t$,
which we denote by $\psi_{p}(t)$.
The equation for $\psi_p(t)$ is expressed as
\begin{equation}
\left[\hat L_t
+(p^2+9/4) H^2\right]
   \psi_{p}(t)=0.
\end{equation}
Solving this equation, we obtain
\begin{eqnarray}
\psi_p(t) & = &
    \sqrt{1\over 2\pi} e^{(-ip-3/2)Ht}.
\label{timefun}
\end{eqnarray}
Note that
$(p^2+9/4)H^2$ is the four-dimensional effective mass squared
for each mode.
The functions $\psi_p(t)$ satisfy the orthonormality
and the completeness conditions,
\begin{equation}
 \int_{-\infty}^{\infty} dt\, H e^{3Ht}
   \psi_p(t)\psi_{p'}^*(t) =\delta (p-p'),
\quad \quad
\int_{-\infty}^{\infty} dp\,
   \psi_p(t)\psi_p^*(t') ={\delta (t-t')\over H e^{3Ht}}\,.
\end{equation}

Next we consider the eigenfunctions of $\hat L_r$, which
we denote by $u_p(r)$. The equation for $u_p(r)$ is
\begin{equation}
 \left[\hat L_r +{p^2+9/4\over \ell^2\sinh^2(r/\ell)}\right] u_p(r)=0.
\end{equation}
We denote the eigenfunction which is regular
on the upper half complex $p$ plane
by $u_p^{\mathrm{(out)}}(r)$, which is given by
\begin{eqnarray}
 u_p^{\mathrm{(out)}}(r) & = &
  {\Gamma(1-ip)\over2^{ip}}{P_{\nu-1/2}^{ip}(\cosh (r/\ell))
  \over{\sinh^{3/2}(r/\ell)}} \cr
   &=&  [\sinh(r/\ell)]^{-ip-3/2}
     [\cosh(r/\ell)]^{\nu+ip-1/2} F\left({-ip-\nu+3/2\over 2},
      {-ip-\nu+1/2\over 2}, -ip+1;\tanh^2(r/\ell)\right)\cr
    &\displaystyle\mathop{\sim}_{r\to 0} & (r/\ell)^{-ip -3/2},
\end{eqnarray}
where $P_{\nu-1/2}^{ip}(z)$ is the associated Legendre function of the
first kind \cite{bateman} and
\begin{equation}
 \nu=\sqrt{m^2\ell^2+4}.
\label{defnu}
\end{equation}
The reason for assigning the superscript `(out)' to this eigenfunction
is that $\psi(t) u_p^{\mathrm{(out)}}(r)\propto e^{-ip(Ht+\ln r)}$
describes a wave propagating out to the Cauchy horizon
given by $r\to0$ with $Ht+\ln r=\mbox{const}$. 
On the other hand,
we denote the eigenfunction which satisfies the Neuman boundary condition
at the position of the brane (\ref{bc}) by $u_p^{(Z_2)}(r)$.
Here the superscript `$(Z_2)$' is assigned because of its $Z_2$-symmetric
property.
We can describe $u_p^{(Z_2)}(r)$ by a linear combination
of two independent solutions as
\begin{equation}
  u_p^{(Z_2)}(r)=u_p^{\mathrm{(out)}}(r)-\gamma_p u_{-p}^{\mathrm{(out)}}(r).
\end{equation}
The Neuman boundary condition determines $\gamma_p$ as
\begin{equation}
 \gamma_p= \left.{\partial_r u_p^{\mathrm{(out)}}(r)\over
        \partial_r u_{-p}^{\mathrm{(out)}}(r)}\right\vert_{r=r_0}.
\label{pole}
\end{equation}

With these eigenfunctions, we can
express the Green function as
\begin{equation}
 {\cal G}(t,r;t',r')
   =\int_{-\infty}^\infty dp \,{\cal G}_p(r,r')\psi_p(t)\psi_{-p}(t'),
\label{calG2}
\end{equation}
where ${\cal G}_p(r,r')$ is given by
\begin{equation}
  {\cal G}_p(r,r')={1\over W_p}
    \left(u_p^{\mathrm{(out)}}(r) u_p^{(Z_2)}(r') \theta(r'-r)
          +u_p^{\mathrm{(out)}}(r') u_p^{(Z_2)}(r) \theta(r-r')\right),
\label{gp}
\end{equation}
and $W_p$ is the Wronskian defined by
\begin{eqnarray}
 W_p & \equiv & \ell^4 \sinh^4(r/\ell)
       \left[ (\partial_r u_p^{\mathrm{(out)}}(r))  u_p^{(Z_2)}(r)
              -u_p^{\mathrm{(out)}}(r) (\partial_r u_p^{(Z_2)}(r))\right]\cr
     & = & 2ip \ell^3 \gamma_p.
\label{wronskian}
\end{eqnarray}
It is straightforward to show that ${\cal G}_{p}(r,r')$ satisfies
\begin{equation}
-\left[\hat L_r+{p^2+9/4\over \ell^2 \sinh^2(r/\ell)}\right]
{\cal G}_p(r,r')= {\delta(r-r')\over \ell^4 \sinh^4(r/\ell)}\,,
\label{Gpeq}
\end{equation}
and that ${\cal G}(t,r;t',r')$ given by Eq.~(\ref{calG2})
satisfies Eq.~(\ref{green2}).

In the above, we have tacitly assumed that there is no pole in the
upper half complex $p$ plane in the Green function ${\cal G}_{p}$.
If this is the case, since the integrand in Eq.~(\ref{calG2}) goes
to zero sufficiently fast at large $|p|$ on the upper half complex
$p$ plane for $t<t'$, the retarded boundary condition
is guaranteed by closing the integration contour in the upper half
complex $p$ plane. In the case when there exist poles in the upper half
complex $p$ plane, one has to deform the integration contour so that
there is no pole inside the contour. In other words, one has to
subtract the contribution of the poles in the upper half complex $p$ plane
from the Green function (\ref{calG2}).
Thus the correct expression for the retarded Green function is
\begin{equation}
 {\cal G}(t,r;t',r')
   =\int_{-\infty}^\infty dp \,{\cal G}_p(r,r')\psi_p(t)\psi_{-p}(t')
-2\pi i\sum_{i}\mathop{\mbox{Res}}_{p=p_i}
\left[{\cal G}_p(r,r')\psi_p(t)\psi_{-p}(t')\right],
\label{calG2c}
\end{equation}
where $p=p_i$ are the locations of poles in the upper half complex $p$ plane.

Once the retarded boundary condition is satisfied, the Green function
vanishes for spatially separated $(t,r)$ and $(t',r')$.
In particular, it vanishes in the limit $r\to 0$ for fixed values
of $t$ and $t'$. This guarantees the
regularity at $r=0$ of the scalar field for arbitrary, regular
initial data.

On the other hand, when $(t,r)$ is in the causal future of $(t',r')$,
one can close the integration contour on the lower half of the complex
$p$ plane. Then the late-time behavior is understood by
investigating the structure of
singularities such as poles and branch cuts
in the integrand.
The singularity on the complex $p$ plane with
the largest imaginary part
dominates the late-time behavior.
In the next section, we investigate the structure of
the singularities in detail.

\subsection{Poles of the Green function}

Let us investigate the singularities in ${\cal G}_p$.
{}From Eq.~(\ref{gp}), the location of a pole
is determined by the value of $p$ at which the Wronskian
(\ref{wronskian}) vanishes.
Noting Eq.~(\ref{pole}),
the equation to determine the pole location is given by
\begin{equation}
{\partial_{r} u_p^{\mathrm{(out)}}(r)}|_{r=r_{0}}=0.
\label{eigeneq}
\end{equation}
We look for the solution $p$ satisfying Eq.~(\ref{eigeneq})
when $H^{2}\ell^{2}\ll 1$ and $m^{2}\ell^{2}\ll 1$.
For this purpose, it is convenient to rewrite the
expression for $ u_p^{\mathrm{(out)}}(r)$ as\cite{bateman}
\begin{eqnarray}
 u_p^{\mathrm{(out)}}(r) 
& = & {\Gamma(1-ip)\over2^{ip}\sqrt{\pi}}
     \Biggl[
            {2^{\nu-1/2}\Gamma(\nu)\over \Gamma(-ip+\nu+1/2)}
           \sinh^{\nu-2}(r/\ell)
            F\left({ip-\nu+1/2\over 2},
      {-ip-\nu+1/2\over 2}, -\nu+1;{-1\over\sinh^2(r/\ell)}\right)
\cr
    & &\quad+
{2^{-\nu-1/2}\Gamma(-\nu)\over \Gamma(-ip-\nu+1/2)}
           \sinh^{-\nu-2} (r/\ell)
            F\left({ip+\nu+1/2\over 2},
            {-ip+\nu+1/2\over 2}, \nu+1;{-1\over\sinh^2(r/\ell)}\right)
\Biggr].
\label{hyper}
\end{eqnarray}
Taking $1/\sinh^{2}(r_0/\ell)=H^{2}\ell^{2}$ as a small parameter,
we expand Eq.~(\ref{eigeneq}).
Besides an irrelevant common factor, we obtain
\begin{eqnarray}
&& {4(1-\nu)(2-\nu)\over 4-\nu}
-\left(p^{2}+\left(-\nu+{1\over2}\right)^{2}\right)H^{2}\ell^{2}
+  {(p^{2}+\left(-\nu+{1\over2}\right)^{2})(p^{2}+(-\nu+{5\over2})^{2})
\over 8(2-\nu)(4-\nu)}H^{4}\ell^{4}
+O\left({p^{6}H^{6}\ell^{6}\over 2-\nu}\right) \cr
&&\quad + {2^{-2\nu+2}(1-\nu)(\nu+2)  \Gamma(-\nu)\Gamma(-ip+\nu+{1\over2})
\over (4-\nu)\Gamma(\nu)\Gamma(-ip-\nu+1/2)} \,
      \,   (H\ell)^{2\nu} \Bigl[ 1 + O(p^{2}H^{2}\ell^{2})\Bigr] =0.
\label{equation1}
\end{eqnarray}
Next we expand this equation with respect to $m^{2}\ell^{2}$.
Then, we notice that the third and all the higher-
order terms on the first line diverge at $m^2\ell^2=0$
because $\nu \approx 2+m^{2}\ell^{2}/4$.
The same is true for the terms on the second line.
However, it is guaranteed that
the singular pieces of these terms mutually cancel
because $P_{3/2}^{ip}(\cosh(r_{0}/\ell))$ is finite.
After the subtraction of these singular pieces,
the terms on the second line are at most
$O(H^{4}\ell^{4})$ except for the cases in which
the argument of the $\Gamma$ function in
the numerator, $-ip+\nu+1/2$, is very close to a nonpositive integer.
Except for these special cases,
the first two terms on the first line
give an approximate equation to determine the location of a pole:
\begin{equation}
{m^{2}\ell^{2}\over 2}
-\left(p^{2}+{9\over4}\right)H^{2}\ell^{2} \approx 0.
\label{leading}
\end{equation}
This equation is always a good approximation irrespective of the value of
$m^{2}/H^{2}$. We denote the two solutions by
\begin{equation}
 p_{\pm} =\pm i\sqrt{{9\over 4}-{m_{\mathrm{eff}}^2\over H^2}}
+O\left((H\ell)^2,(m\ell)^2\right),
\label{ppm}
\end{equation}
where
\begin{equation}
m_{\mathrm{eff}}^{2}
={m^{2}\over 2}.
\label{mass}
\end{equation}
In addition to these, there are poles corresponding to the
poles of the $\Gamma$ function mentioned above,
given by $p\approx -i(n+\nu+1/2)$, where $n$ is a non-negative integer.
However, they do not give a dominant contribution to
the late-time behavior.

Now it is easy to see the late-time behavior of the Green function.
When $m_{\mathrm{eff}}^{2}/H^{2}<9/4$,
the contribution from the pole $p_+$ dominates.
The Green function after a sufficiently long lapse of time behaves as
\begin{equation}
 {\cal G}(x,x') \propto e^{(\sqrt{(9/4)H^2-m_{\mathrm{eff}}^2}-(3/2)H)t}.
\label{asym1}
\end{equation}
When $m_{\mathrm{eff}}^{2}/H^{2}>9/4$,
the contributions from both poles $p_\pm$ are equally important.
In this case, the asymptotic behavior of the Green function is given by
\begin{equation}
 {\cal G}(x,x') \propto e^{-(3/2)H t}
   \cos\left(
   \left[{m_{\mathrm{eff}}^2-{9\over 4}H^2}\right]^{1/2}t+\eta
\right),
\label{asym2}
\end{equation}
where $\eta$ is a real constant phase.

According to Dubovsky et al.\cite{Dubovsky:2000am},
a massive bulk scalar field in the Randall-Sundrum braneworld with a
single flat brane has quasinormal mode poles at
$p\approx \pm m/\sqrt{2}-i\epsilon$,
where $\epsilon$ is a small
positive number, which correspond to a resonant state with the
effective mass squared $m^2/2$.
In addition to these, as is explained in detail in Appendix \ref{flat},
there also exists a branch cut emanating from
$p^2=0$ in the flat brane case, though its contribution is
physically small compared with the contribution from the pole.
There is of course a nice correspondence between these
properties of the singularities in the flat case and
those in the present de Sitter case.
The quasinormal mode poles are merely the poles at $p=p_{\pm}$
discussed above.
At the leading order the location of these poles is determined by
Eq.~(\ref{leading}). But it is not sufficient
to identify the small imaginary part in $p^2_{\pm}$.
To do so, we have to go back to Eq.~(\ref{equation1}).
Then we find the higher-order
terms in the first line of Eq.~(\ref{equation1})
are irrelevant to obtain the imaginary part
since they are real functions of $p^2$,
but the essential term is the first term in the second line of
Eq.~(\ref{leading}).  Taking that term into account,
the leading-order contribution to the imaginary part
is obtained as
\begin{equation}
{H^2\ell^2} \mbox{Im}\left[p_{\pm}^{2}\right]
\approx
\mbox{Im}\left[{2^{-2\nu+2}(1-\nu)(\nu+2)
\Gamma(-\nu)\Gamma(-ip_{\pm}+\nu+1/2)
\over (4-\nu)\Gamma(\nu)\Gamma(-ip_{\pm}-\nu+1/2)} \,
(H\ell)^{2\nu}\right]
\mathop{\longrightarrow}\limits_{m^2/H^2\to\infty}
 -{(m\ell)^{4}\over 16}\pi\,.
\end{equation}
This result completely agrees with that obtained in \cite{Dubovsky:2000am}.
As for the branch cut in the flat case, it corresponds to
the series of the poles at $p\approx-i(n+\nu+1/2)$
in the limit $H/m\to0$.

\section{Effective four-dimensional scalar field}

{}From the asymptotic behavior obtained in
Eqs.~(\ref{asym1}) and (\ref{asym2}), we can conclude that the bulk
scalar field evaluated on the brane behaves as an effective
four-dimensional field with the mass squared $m_{\mathrm{eff}}^{2}$
after a sufficiently long period of de Sitter expansion.
This correspondence was speculated in
Refs.~\cite{Himemoto:2001nd,Sago:2001gi}
based on an analysis of the field configuration
in the separable form as Eq.~(\ref{sepform})
for the case $|m^2|/H^2\ll1$.
Our present analysis proves that
this simple correspondence continues to hold
in more general cases, irrespective of initial field
configurations and of the value of $m^2/H^2$.

This is not the whole story of the correspondence.
Our analysis developed so far,
in which we fixed the background geometry to a model
with a single de Sitter brane,
can be interpreted as a first-order approximation with
respect to the amplitude of the scalar field.
{}From Eq.~(\ref{friedmann}), we see that
the backreaction to the geometry starts with the second order
in the amplitude of $\phi$.
Thus, it is justified to evaluate
the effective energy density $\rho_{\mathrm{eff}}$ defined in
Eq.~(\ref{rhoeff}) by
substituting the leading-order evolution of the scalar field
on the fixed background.
It is then important to check whether $\rho_{\mathrm{eff}}$,
which describes the effect of the backreaction to the geometry,
coincides with the energy density of the effective four-dimensional
field with $m^2_{\mathrm{eff}}=m^2/2$.
In Ref.~\cite{Sago:2001gi}, this was shown to be the case
for $|m^2|/H^2\ll 1$ and for the scalar
field configuration in the separable form.
Here we shall show that this correspondence also continues to hold
in general, irrespective of the form of the scalar field
configuration and of the ratio $m^2/H^2$.

To see this correspondence, we first note that
from the asymptotic behavior of the Green function
(\ref{asym1}) and (\ref{asym2}), 
the bulk scalar field satisfies the equation
\begin{eqnarray}
\ddot\phi+3H\dot\phi+{1\over2}V'=0
\end{eqnarray}
on the brane at late times.
On the other hand, the five-dimensional field equation (\ref{fieldequ})
implies
\begin{eqnarray}
\ddot\phi+3H\dot\phi-\partial_r^2\phi+V'=0,
\end{eqnarray}
on the brane. Thus we have
\begin{eqnarray}
\partial_r^2\phi={1\over2}V'=-\ddot\phi-3H\dot\phi\,.
\label{dr2phi}
\end{eqnarray}
Inserting this into the integrand of Eq.~(\ref{weyl}),
we obtain
\begin{eqnarray}
E_{tt}=-{\kappa_5^{2}\over{2a^{4}}}\int^t a^{4}\dot{\phi}
(\ddot\phi + 2H\dot{\phi})\,dt
=-{\kappa_5^{2}\over{4a^{4}}}\int^t
{d\over dt}\left(a^{4}\dot{\phi}^2\right)\,dt
=-{\kappa_5^2\over4}\dot\phi^2,
\end{eqnarray}
where we have neglected the integration constant term
($\propto a^{-4}$) that vanishes rapidly as time goes on.
Thus we find $\rho_{\mathrm{eff}}$ is given by
\begin{eqnarray}
\rho_{\mathrm{eff}}={3\over{\kappa_5^{2}\sigma}}
\left({\dot{\phi}^{2} \over 2}+V(\phi)\right)-{E_{tt}\over\kappa_4^2}
={1\over 2}\dot\Phi^2+V_{\mathrm{eff}}(\Phi),
\end{eqnarray}
where
\begin{equation}
 \Phi=\sqrt{\ell_0}\,\phi\,;\quad\ell_0={6\over\kappa_5^2\sigma}\,,
\label{cor1}
\end{equation}
and
\begin{equation}
 V_{\mathrm{eff}}(\Phi)={\ell_0\over 2}V(\Phi/\sqrt{\ell_0}).
\label{cor2}
\end{equation}
It should be noted that not only the mass term
but also the constant vacuum energy term is included
in this scaling relation of the
potentials.

This result is quite suggestive.
Although our analysis is done for the specific form
of the potential (\ref{potential})
on the fixed AdS bulk plus de Sitter brane background,
it is tempting to conjecture that
the relations (\ref{cor1}) and (\ref{cor2}) continue to hold
in more general situations, as far as the dynamics
in the zeroth order
of $H^2\ell^2$ and $m^2\ell^2$ is concerned.
The crucial point in the above arguments was
the equality
\begin{eqnarray}
\partial_r^2\phi={1\over2}V'(\phi)\,,
\end{eqnarray} 
on the brane. In other words, if this equality holds
at the zeroth order of $H^2\ell^2$ and $m^2\ell^2$,
the effective dynamics of the Einstein-scalar system
on the brane will be almost identical to the
corresponding system in the standard four-dimensional theory.

\section{summary and discussion}

We have analyzed the late-time behavior of a bulk scalar
field in the inflating braneworld of the
Randall-Sundrum type.
We have considered a simple model, in which the background
spacetime is given by the five-dimensional anti-de Sitter space
and the brane is expanding with a constant Hubble rate $H$.
Assuming the spatial homogeneity of the brane,
we have formally solved the field equation of the bulk scalar field
as an initial value problem by using the Green function.

Focusing on the integral representation
of the Green function given in Eq.~(\ref{calG2}),
we have derived an expression for the integrand ${\cal G}_p$.
By analyzing the singularity structure of ${\cal G}_p$
as a function of complex $p$,
we have derived the late-time behavior of the bulk scalar field
under the assumptions $m^{2}\ell^{2}\ll 1$ and $H^{2}\ell^{2} \ll 1$,
where $\ell$ is the curvature radius of AdS$_5$ and
$m$ is the mass of the bulk scalar field.
When $m^2/H^2$ is small, we have found that a single pole in
${\cal G}_p$ dominates the late-time dynamics of the scalar field,
whereas a complex conjugate pair of poles
dominates it when $m^2/H^2$ is large.
The critical value for $m^2/H^2$ is $\approx 9/2$.
Irrespective of the ratio $m^{2}/H^{2}$, the
bulk scalar field seen on the brane behaves as
a four-dimensional effective scalar field with mass
$m_{\mathrm{eff}}=m/\sqrt{2}$.

Using the late-time behavior of the bulk scalar field $\phi$
as described above, we have examined the lowest-order
backreaction to the geometry which starts at the
quadratic order in the amplitude of $\phi$.
We have found that the leading-order backreaction to the geometry
is equivalently represented by a four-dimensional effective scalar field
$\Phi$ with the effective four-dimensional mass $m_{\mathrm{eff}}$
mentioned above, where $\Phi$
is related to $\phi$ by a simple scaling (\ref{cor1}).

Although we have studied only the case with
a quadratic potential (\ref{potential}),
we have described this correspondence in
a more suggestive manner,
$V_{\mathrm{eff}}(\Phi)=\ell_0V(\Phi/\sqrt{\ell_0})/2$.
Note that this relation holds not only for the mass term
but also for the vacuum energy term.
We have conjectured that this correspondence
may hold in more general cases. If this is indeed the case,
the braneworld inflation driven by
a bulk scalar field proceeds just like the standard
inflation in the four-dimensional theory, apart from
small corrections of $O(H^2\ell^2,\,m^2\ell^2)$.

Finally, let us make a comment on the effect of the
presence of the imaginary part in the dominant poles $p_{\pm}$.
When we consider the oscillating phase around $\phi=0$,
the existence of the imaginary part causes damping
of the oscillation amplitude.
If we assume $m^2\gg H^2$, the amplitude decays as
\begin{equation}
 |\phi|^2\propto e^{-(3H+\pi m_{\mathrm{eff}}^3 \ell^2/4) t}.
\end{equation}
This means that the energy density of the scalar field oscillation
decreases faster than the rate expected from the cosmic expansion.
This effect, although it seems small, may play some role in
the reheating after inflation.
We plan to come back to this issue in a future publication.

\acknowledgments
We would like to thank J. Yokoyama for useful comments
and discussions.
This work was supported in part by the Monbukagakusho Grant-in-Aid
Nos.~1270154 and 12640269, 
and by the Yamada Science Foundation.

\appendix

\section{Bulk scalar field in the Randall-Sundrum braneworld}
\label{flat}
In this appendix, we consider the behavior of a bulk scalar field
in the Randall-Sundrum braneworld, i.e., the AdS$_5$ bulk
with a flat positive tension brane.
This is just the flat brane limit ($H^2\ell^2\to0$) of the
situation discussed in the text. However, it is worth
comparing both cases because there appears a different aspect in the
late-time behavior in the flat brane limit.
Most of the results in this appendix are a 
rephrasing of those in Ref.~\cite{Dubovsky:2000am}.

In the Randall-Sundrum braneworld, the background metric is given by
\begin{equation}
 ds^2={\ell^2\over \tilde z^2}\left[dz^2-dt^2+d{\bbox{x}}^2\right],
\end{equation}
where $\tilde z = |z|+\ell$.
The brane is located at $z=0$, i.e., $\tilde z=\ell$.
Substituting $\phi=\tilde z^{3/2} e^{-i \omega t} v_{\omega}(z)$
into the field equation,
we obtain the equation for the
$z$-dependent part of the mode function as
\begin{equation}
 \left[\partial_z^2+\omega^2-\tilde z^{-2}
     \left(m^2\ell^2+{15\over 4}\right)+
     {3\over \ell}\delta(z)\right]
     v_{\omega}(z) =0.
\end{equation}
Solving this equation,
we obtain the outgoing wave solution as
\begin{eqnarray}
 v_{\omega}^{\mathrm{(out)}}(z) = \sqrt{\pi\tilde z \omega \over 2}
         H_{\nu}^{(1)}\left(\omega\tilde z\right).
\end{eqnarray}
In the same way as we have done in the main text,
we can construct the Green function as
\begin{equation}
 G = \int_{-\infty}^{\infty} d\omega \,
{\cal G}_{\omega}(z,z'){e^{-i\omega (t-t')}\over 2 \pi},
\label{formalG}
\end{equation}
where
\begin{equation}
{\cal  G}_\omega(z,z')= {1\over W_\omega}
      \left[v_\omega^{\mathrm{(out)}}(z) v_\omega^{(Z_2)}(z')
          \theta(|z|-|z'|)
       +v_\omega^{(Z_2)}(z) v_\omega^{\mathrm{(out)}}(z')
          \theta(| z'|-|z|)\right],
\label{Gw}
\end{equation}
with
\begin{equation}
 W_\omega= (\partial_z v_\omega^{\mathrm{(out)}}) v_\omega^{(Z_2)}
               -v_\omega^{\mathrm{(out)}} (\partial_z v_\omega^{(Z_2)}).
\end{equation}

Then, the equation which determines the location of the poles
of ${\cal G}_{\omega}$ is given by
\begin{equation}
\partial_{z}v_{\omega}^{\mathrm{(out)}}(z)|_{z=0}=0.
\end{equation}
As we are interested in the late-time behavior,
which is expected to be dominated by $\omega$ with
$\omega \ell \ll 1$ when $m^{2}\ell^{2}\ll 1$,
we expand the above equation around $\omega=0$ as
\begin{equation}
  \omega^2-{m^2 \over 2}+i\omega^4\ell^2
    \left({\pi\over 4}-{\mbox{arg}(\omega) \over 2}\right)\approx 0.
\end{equation}
As a result, we find the quasinormal mode poles
at\cite{Dubovsky:2000am}
\begin{equation}
 \omega_{\pm p}\approx \pm {m\over \sqrt{2}}
    -{\pi m^{3}\ell^2\over 16\sqrt{2}}i\,.
\end{equation}

In addition to the above poles,
there exists a brunch cut emanating from $\omega=0$.
To evaluate this contribution,
we expand the integrand of Eq.~(\ref{formalG}) assuming that
$\tilde z\approx \ell$,
$\omega \tilde z\ll 1$ and $\tilde z <\tilde z'$.
Then, we find
\begin{eqnarray}
G \approx
{i \ell\over 4} \left({\tilde z\over \ell}\right)^{-\nu+(1/2)}
\left({\tilde z'\over \ell}\right)^{1/2}
\int_{-\infty}^{\infty} d\omega\,
           {e^{-i\omega(t-t')} \, \omega^\nu \,
H_{\nu}^{(1)}(\omega\tilde z')
          \over \omega^2-{m^2\over 2}+i\omega^4\ell^2
           \left({\pi\over 4}-{\mbox{arg}(\omega) \over 2}\right)},
\label{GF}
\end{eqnarray}
where $\nu=\sqrt{4+m^2\ell^2}\approx 2$.
To proceed the evaluation of the contribution from the branch cut,
we perform a modification of the integration contour.
Noting that
\begin{eqnarray}
  H_{\nu}^{(1)}( e^{-\pi i/2} s \tilde z')
   &=&{2 e^{-\nu\pi i/2}\over \pi i}
     \left[e^{\nu\pi i} K_\nu(s\tilde z')+\pi i I_{\nu}(s\tilde z')\right],
\cr
  H_{\nu}^{(1)}( e^{3\pi i/2} s \tilde z')
   &=&{2 e^{-\nu\pi i/2}\over \pi i}
     \left[e^{-\nu\pi i} K_\nu(s\tilde z')-\pi i I_{\nu}(s\tilde z')\right],
\end{eqnarray}
we obtain
\begin{eqnarray}
 G_{\mathrm{cut}} \approx
           -\left({\tilde z\over \ell}\right)^{-3/2}
          \left({\tilde z'\over \ell}\right)^{1/2}
         &&
           \int_{0}^{\infty} {ds\over s}
            {e^{-s(t-t')} \over 4}
          {s^4 \over (s^2+{m^2\over 2})^2} \cr
      &&   \times\left[(s\ell)^3 K_{\nu}(s\tilde z')
               +4\ell \left(s+{m^2\over 2s}\right)
               I_{\nu}(s\tilde z')\right].
\label{Gflatcut}
\end{eqnarray}
The behavior of $K_{\nu}(s\tilde z')$ and
$I_{\nu}(s\tilde z')$ is
\begin{eqnarray}
&&K_{\nu}(s\tilde z')\approx
\left\{\begin{array}{ll}
 ({1/ 2}) \left({s\tilde z'/ 2}\right)^{-2}
&\quad\mbox{for~}s\tilde z'\ll1\,,
\\
 ({\pi e^{-s\tilde z'}/\sqrt{2s\tilde z'}})
&\quad\mbox{for~}s\tilde z'\gg1\,,
\end{array}\right.
\nonumber\\
&&I_{\nu}(s\tilde z')\approx
\left\{\begin{array}{ll}
 ({1/ 2}) \left({s\tilde z'/2}\right)^2
&\quad\mbox{for~}s\tilde z'\ll1\,,
\\
({e^{s\tilde z'}/\sqrt{2\pi s\tilde z'}})
&\quad\mbox{for~}s\tilde z'\gg1\,.
\end{array}\right.
\nonumber
\end{eqnarray}
For $m(t-t')\gg1$ and $t-t'\gg\tilde z'$,
the integral in Eq.~(\ref{Gflatcut})
is dominated by the contribution from $s\ll m$.
Therefore, it is estimated as
\begin{equation}
 G_{\mathrm{cut}} \approx
          -\left({\tilde z\over \ell}\right)^{-3/2}
         \left({\tilde z'\over \ell}\right)^{-3/2}
          {4!\,m\ell\over m^{5}(t-t')^5}
       \left[2+{m^{2}\ell^2\over4}
\left({\tilde z'\over \ell}\right)^{4}\right].
\label{cutestimate}
\end{equation}
Thus the contribution
from the branch cut is not exponentially suppressed as is
usually the case, and it dominates over the
pole contribution after the epoch
$t-t'\approx ({\pi m^{3} \ell^2\over 16\sqrt{2}})^{-1}$
when the exponential damping of the
pole contribution becomes significant.
By this epoch, however, the contribution from the branch cut
is already suppressed by a factor
\begin{equation}
\sim \left({\pi m^2\ell^2\over 16\sqrt{2}}\right)^5 m\ell\,.
\end{equation}
Thus, unless the second term in the square brackets of
Eq.~(\ref{cutestimate}) is extremely large,
the branch-cut contribution is always negligible
under our current assumption $m\ell\ll1$.

Finally, let us mention the fact that
the branch cut is absent in the case of the de Sitter brane.
However, this is not a contradiction.
In the case of the de Sitter brane, there is
a sequence of poles at $p\approx -i(n+\nu+1/2)$ where
$n$ is a non-negative integer.
As we can see from the comparison of the time-dependent
part of the mode functions, $\omega$ used in this appendix
corresponds to $Hp$ in the main text.
In terms of $\omega$, the locations of the poles
are at $\omega\approx -i H(n+\nu+1/2)$, and hence
their distribution becomes continuous in the limit
$H\to 0$.

\end{document}